\documentclass[draft]{agujournal2019}
\usepackage{url}
\usepackage{lineno}
\usepackage[finalnew]{trackchanges}
\usepackage{soul}
\usepackage{amsmath,amsfonts,amssymb}

\draftfalse

\journalname{JGR: Space Physics}

\newcommand{\mwm}{mW\,m$^{-2}$}
\newcommand{\keV}{keV}
\newcommand{\MeV}{MeV}
\newcommand{\ratio}{$\Phi_\uparrow / \Phi_\downarrow$}

\begin{document}

\title{Hemispheric asymmetry and up/down anisotropy of Jupiter's auroral electron precipitation: Juno/JEDI statistics}

\authors{B. Benmahi\affil{1,2}, N. Andr\'e\affil{3,4}, V. Hue\affil{1}, Z.-Y. Liu\affil{3}, M. Blanc\affil{5,3}, L. Head\affil{2}, M. Devinat\affil{3}, G. Gronoff\affil{6,7}, D. C. Grodent\affil{2}, and B. Bonfond\affil{2}}

\affiliation{1}{Aix-Marseille Universit\'e, CNRS, CNES, Institut Origines, LAM, Marseille, France}
\affiliation{2}{Laboratory of Atmospheric and Planetary Physics, STAR Institute, University of Li\`ege, Li\`ege, Belgium}
\affiliation{3}{Institut de Recherche en Astrophysique et Plan\'etologie (IRAP), CNES-CNRS-Universit\'e de Toulouse, Toulouse, France}
\affiliation{4}{Institut Sup\'erieur de l'A\'eronautique et de l'Espace (ISAE-SUPAERO), Universit\'e de Toulouse, Toulouse, France}
\affiliation{5}{Weihai Institute for Interdisciplinary Research, Shandong University, Weihai, China}
\affiliation{6}{Science Systems and Applications Inc., Hampton, VA, USA}
\affiliation{7}{NASA Langley Research Center, Hampton, VA, USA}

\correspondingauthor{Bilal Benmahi}{bilal.benmahi@lam.fr}

\begin{keypoints}
\item Electron energy flux into the Jovian aurora differs by about a factor of two between the north and the south
\item The dominant hemisphere is the north in the polar emission and the south in the main emission
\item Mean electron energy is the same in both hemispheres so the asymmetry lies in the electron number flux
\end{keypoints}

\begin{abstract}
Energetic electron precipitation shapes Jupiter's auroral emissions, thermal structure, and atmospheric chemistry. Whether its intensity differs between hemispheres, and whether that difference is consistent across auroral sub-regions, remains statistically unexplored over a large perijove sample. We statistically characterize the precipitating electron energy flux in Jupiter's auroral sub-regions, polar emission (PE) and main emission (ME), separating downward and upward fluxes and comparing hemispheres, using Juno/JEDI data from PJ3 to PJ66. Juno footprints from JRM33 and a current sheet model assign each spectrum to an auroral sub-region, where the downward and upward energy flux are integrated over the corresponding loss cone. The median energy flux differs between hemispheres by a factor of about two in the PE and the ME, for both flux directions, but the dominant hemisphere is reversed, the north dominating in the PE (N/S of 2.5 downward, 2.0 upward) and the south in the ME (0.65 and 0.5). The mean electron energy is nearly identical in both hemispheres in every sub-region, so the asymmetry lies in the number flux rather than the local acceleration efficiency. The PE also shows a systematic upward excess over downward flux in both hemispheres, maximal at the magnetic pole. The PE/ME reversal shows that the two hemispheres partition precipitated energy differently between the polar cap and the main oval. The PE anisotropy, qualitatively identical in both hemispheres, indicates that this partition is not driven by a hemispheric contrast in field-aligned acceleration efficiency.
\end{abstract}

\section*{Plain Language Summary}
Jupiter's auroras are powered by energetic electrons that travel along the planet's magnetic field and strike the upper atmosphere near the poles. These electrons heat the atmosphere and drive chemical reactions. Until now, it was not known whether the northern and southern auroras receive the same amount of electron energy. Using measurements from the Juno spacecraft collected over 64 close passes of Jupiter (perijoves 3 to 66), we compare the electron energy reaching the north and the south, separately for the region inside the main auroral oval, called the polar emission, and for the main oval itself. We find that the north receives about twice as much energy as the south in the polar region, whereas the south receives about twice as much as the north in the main oval. The energy carried by a single electron is the same in both hemispheres, so the difference comes from the number of electrons and not from how strongly they are accelerated. Inside the polar region, electrons traveling away from Jupiter carry more energy than those traveling toward it, in both hemispheres. These results show that the two hemispheres share the incoming energy differently, and that models of Jupiter's upper atmosphere should not use the same electron input in the north and in the south.

\section{Introduction}
\label{sec:intro}

Jupiter's magnetic field guides energetic electrons from the magnetosphere toward the northern and southern polar regions, where they give rise to the most powerful planetary auroral emissions in the solar system \cite{Broadfoot1979,Clarke1980}. The auroral emission regions can be divided into sub-regions in several ways, depending on the study considered. In this work, the aurora in each hemisphere is divided into the main emission (ME), the polar emission (PE) located poleward of it, and the outer emission (OE) located equatorward of it and which is not considered in this work \cite{Grodent2015,Benmahi2024a,Benmahi2024b,Vinesse2026}.

In addition to producing these emissions, this energy input heats the auroral upper atmosphere and modifies its chemistry \cite{Sinclair2018,Cavalie2023,Hue2024}. Juno, in polar orbit around Jupiter since 2016 \cite{bolton_juno_2017}, provided the first in situ measurements of the energy and angular distributions of precipitating electrons, using the Jovian Auroral Distributions Experiment (JADE; \citeNP{McComas2017}, $\sim$0.1--100\,\keV{}) and the Jupiter Energetic particle Detector Instrument (JEDI; \citeNP{Mauk2017_instrument}, $\sim$30\,\keV{}--1\,\MeV{}). These observations revealed that electrons above the polar cap also travel persistently upward. \citeA{Mauk2017_beams} observed, from the first perijove, uni- or bidirectional field-aligned beams with unstructured spectra in both hemispheres, and \citeA{Allegrini2017} confirmed this at lower energies, further detecting upward beams and upward-pointing loss cones at latitudes magnetically connected to the diffuse aurora \cite<e.g.,>{li_understanding_2017}.

\citeA{Mauk2020} reviewed the diversity of mechanisms driving downward precipitation above the ME, from pitch-angle scattering of a trapped population at the lowest auroral latitudes to field-aligned acceleration by a parallel potential drop at higher auroral latitudes, producing quasi-monoenergetic (\textit{inverted-V}) or power-law (\textit{broadband}) spectra in varying proportions \cite{Mauk2018,Salveter2022}. Closer to the polar cap, a third zone displays bidirectional acceleration, where upward beams add to and can locally exceed the downward precipitation.

Jupiter's internal magnetic field is itself hemispherically asymmetric. Its non-dipolar component is concentrated in a narrow northern band, with part of the flux returning through an isolated, intense patch near the equator \cite{Moore2018}, while the south remains predominantly dipolar. This asymmetry is reflected in the kidney-bean shape of the northern ME \cite{Grodent2008a}, which significantly differs from the more regular oval-like southern ME, and is accounted for by the JRM33 model \cite{Connerney2022}. Such asymmetries have long been neglected in magnetosphere-ionosphere coupling models \cite{Bonfond2020}, yet \citeA{Kotsiaros2019} report field-aligned currents approximately twice as intense in the south, attributed to a north-south difference in ionospheric conductance for which \citeA{Sicorello2025} provide tentative case-study evidence. Comparisons of the auroral brightness between the two hemispheres have been carried out for specific structures observed quasi-simultaneously \cite{gerard_hubble_2013,bonfond_evolution_2013,bonfond_dynamics_2016,head_characterising_2026}, but whether the precipitating electron energy flux itself exhibits a comparable hemispheric contrast, and whether that contrast is the same across the auroral sub-regions, has not been assessed statistically over a large number of perijoves.

In this paper, we use JEDI data from PJ3 to PJ66 to statistically characterize the precipitating electron energy flux above 30 keV in the auroral ME and PE sub-regions, separating the upward and downward components, and to quantify how they differ between the northern and southern hemispheres.

\section{Data and method}
\label{sec:data}

\subsection{Instrument and data processing}
\label{sec:instrument}

This study relies on data from the Jupiter Energetic particle Detector Instrument (JEDI; \citeNP{Mauk2017_instrument}) acquired from PJ3 to PJ66. This instrument is equipped with three solid-state detector heads that measure electron energies from $\sim$30\,\keV{} to $\sim$1\,\MeV{} with nearly complete instantaneous pitch-angle coverage and a cadence of $\sim$0.5\,s, averaged to 1\,s.

We use level-3 (L3) differential electron flux data from JEDI, downloaded from the PDS database \cite{Mauk2026}. For each spectrum, these data provide the electron differential intensity as a function of energy, time and pitch angle. Two corrections are applied prior to the statistical analysis. From PJ20 onward, a known processing error affects the small pixel data below $\sim$50\,\keV{}. This error has since been corrected in the most recent JEDI calibration matrices, but it is still present in the archived data used here (B. Mauk, private communication). The affected portion is replaced by a power-law extrapolation, with the spectral index individually fitted for each spectrum by a least-squares regression in log-log space over the [50\,\keV{}; 120\,\keV{}] range. The Minimum Ionizing Bump, a spurious peak centered at $\sim$160\,\keV{} caused by energetic electrons penetrating the solid-state detectors themselves \cite{Mauk2017_instrument,Mauk2026}, is removed by fitting the spectral continuum between $\sim$100 and $\sim$280\,\keV{} with a smoothing function, then subtracting the extrapolated component beneath the peak. The electron dataset resulting from these corrections is the same as that used in \citeA{benmahi_precipitation_2026}, which compares the precipitating energy flux of protons and electrons in the same auroral sub-regions.

To further limit contamination by penetrating radiation from Jupiter's radiation belts, only measurements acquired at a planetocentric distance below 200\,000\,km ($\sim$3\,R$_J$) are included in the statistical analysis. Table~\ref{tab:sample_sizes} reports the resulting number of grid-cell averaged spectra (Section~\ref{sec:mapping}) retained in each auroral sub-region, hemisphere, and flux direction.

The disturbances due to penetrating radiation background are not unique to JEDI. The same contamination signature, an increase in apparent electron flux near the low-latitude boundary of the OE not matched by a corresponding increase in UV auroral brightness, is also present in JADE electron measurements over the same region. For JEDI, this penetrating background near Jupiter's radiation belts is dominated by protons above $\sim$2.5\,GeV rather than by energetic electrons \cite{kollmann_jupiters_2021}, which makes it difficult to separate from the genuine precipitating electron signal using energy or pitch-angle criteria alone. No method currently exists to reliably subtract this contribution, owing to the difficulty of modeling the penetrating background quantitatively (B. Mauk, private communication).

\begin{table}[h]
\caption{Number of grid-cell averaged JEDI spectra retained in each auroral sub-region, hemisphere, and flux direction, after the planetocentric distance selection described above.}
\centering
\begin{tabular}{lcccc}
\hline
& \multicolumn{2}{c}{Downward} & \multicolumn{2}{c}{Upward} \\
& North & South & North & South \\
\hline
PE & 3819 & 1005 & 5083 & 5441 \\
ME & 2123 & 760  & 2024 & 802  \\
\hline
\end{tabular}
\label{tab:sample_sizes}
\end{table}

\subsection{Magnetic mapping and assignment to sub-regions}
\label{sec:mapping}

At each time step, Juno's position is computed from the mission's SPICE kernels, then traced along the local magnetic field line down to Jupiter's atmosphere using the JRM33 magnetic field model combined with the CON2020 current sheet model \cite{Connerney2020,Connerney2022}, yielding the spacecraft's magnetic footprint in System III planetocentric coordinates. The field line is traced to the northern hemisphere when Juno is north of the equator, and to the southern hemisphere otherwise. Owing to Juno's orbital geometry, the northern and southern measurements compared in this study are not simultaneous. Successive footprints are sampled on a $0.33^\circ \times 0.33^\circ$ latitude-longitude grid ($\sim$400\,km in latitude at the reference auroral altitude of 400\,km, Section~\ref{sec:fluxes}), and an average spectrum is computed in each cell crossed. These grid-cell averaged spectra, rather than the individual 1\,s spectra, are the statistical units used throughout this study.

Each measured electron distribution is assigned to one of the three auroral sub-regions, in either the northern or southern hemisphere. The ME region is bounded by contours derived from Juno/UVS observations for each perijove \cite{Groulard2024}, the same masks as those used in \citeA{Benmahi2024a,Benmahi2024b} and \citeA{Vinesse2026}. Perijove-specific masks are used whenever available, since the main emission is known to vary in size \cite{head_effect_2024}. These masks are available up to PJ39, so for PJ40 to PJ66 the ME is bounded by the mean contour of the PJ1 to PJ39 masks. The PE region corresponds to the area inside the inner boundary of the ME, and the OE region to the area between the outer boundary of the ME and the mean magnetic footprint of Io, determined using the JRM33 model combined with a current sheet model \cite{Connerney2020,Connerney2022}. Data acquired when Juno's magnetic footprint is located outside these three sub-regions are excluded from the analysis. We note that the precipitating energy flux increases markedly as the magnetic footprint approaches the outer boundary of the OE region, near the mean magnetic footprint of Io. This enhancement is not associated with a corresponding increase in UV auroral brightness and is therefore unlikely to reflect a genuine increase in precipitating electron flux. It more plausibly results from contamination by penetrating radiation, a known effect for JEDI near the radiation belts \cite{paranicas_radiation_2017}, which the planetocentric distance selection described in Section~\ref{sec:instrument} reduces but does not entirely remove.

\subsection{Computation of downward and upward fluxes}
\label{sec:fluxes}

For each JEDI energy spectrum, the local loss-cone angle $\alpha_\mathrm{LC}$ is computed from the magnetic field evaluated, at each time step, along the traced field line at the spacecraft's position, $B_\mathrm{sc}$, and at the atmospheric level, $B_\mathrm{atm}$, taken at 400\,km altitude corresponding to the mean altitude of the UV auroral emissions \cite{bonfond_far-ultraviolet_2015}, through $\sin^2\alpha_\mathrm{LC} = B_\mathrm{sc}/B_\mathrm{atm}$. Because the direction of Jupiter's internal magnetic field is reversed between the two hemispheres, the precipitating (downward, planetward) electron population corresponds to pitch angles close to $0^\circ$ in one hemisphere and close to $180^\circ$ in the other. We symmetrically define the upward (anti-planetward) population as occupying the opposite pitch-angle range. The downward $\Phi_\downarrow$ and upward $\Phi_\uparrow$ energy fluxes, in \mwm{}, are computed separately by integration over the JEDI energy range:

\begin{linenomath*}
\begin{equation}
\Phi_{\downarrow,\uparrow} = \int_{E_\mathrm{min}}^{E_\mathrm{max}} E \cdot j_{\downarrow,\uparrow}(E)\,\mathrm{d}E,
\label{eq:flux}
\end{equation}
\end{linenomath*}
\noindent where $j_{\downarrow,\uparrow}(E) = \pi\, h_{\downarrow,\uparrow}(E)$ is the differential energy flux distribution obtained from the mean directional differential intensity $h_{\downarrow,\uparrow}(E)$ in the downward or upward loss cone, under the assumption of an isotropic electron distribution within each loss cone, and $[E_\mathrm{min}; E_\mathrm{max}]$ = [30\,\keV{}; 1\,\MeV{}] corresponds to the nominal JEDI energy range.

Since JEDI's two sensor heads viewing within the spacecraft equatorial plane sample a nearly complete pitch-angle distribution almost instantaneously \cite{Mauk2017_instrument}, the downward and upward mean directional intensities are derived simultaneously, from the same $\sim$1\,s measurement, whenever both loss cones are covered. Within each $0.33^\circ \times 0.33^\circ$ grid cell, the resulting $\Phi_\downarrow$ and $\Phi_\uparrow$ values are averaged separately over all contributing time steps before their ratio \ratio{} is computed, so that the comparison is performed cell by cell rather than by contrasting two independently built hemispheric maps.

This study relies solely on JEDI data, without incorporating JADE measurements. The quantity compared throughout this paper is a ratio of median energy fluxes between two hemispheres, so what matters is not the size of the energy range left out but whether leaving it out affects the two hemispheres differently. Below the JEDI threshold of $\sim$30\,\keV{}, the missing part of the spectrum accounts for a median of $\sim$6\% of the downward electron energy flux in both the PE and the ME, a value we obtained by integrating the merged JADE and JEDI spectra of \citeA{benmahi_energy_2026} from PJ3 to PJ35. This proportion is small enough, and close enough between hemispheres, that the omitted range shifts the hemispheric ratios by at most a few percent, without affecting their reversal between the two sub-regions. A second consideration reinforces this choice. Merging the two instruments is only possible over the time steps where both of them sample the same loss cone, which substantially reduces the auroral coverage available for a statistical analysis.

\subsection{Analysis strategy}
\label{sec:strategy}

For the PE and the ME, and for each hemisphere, we construct the statistical distributions of the downward $\Phi_\downarrow$ and upward $\Phi_\uparrow$ energy flux, normalized as a density in logarithmic space ($\mathrm{d}N/\mathrm{d}\log_{10}\Phi$), combining all grid-cell averaged spectra from PJ3--PJ66. These distributions are compared pairwise between the northern and southern hemispheres, for each sub-region and each flux direction, through their medians and the ratio of these medians (Section~\ref{sec:uncertainty}). The ratio \ratio{} is computed in the three sub-regions, including the OE, where the flux distributions themselves are not explored further.

The degree to which one hemisphere dominates the other is quantified by the non-parametric Brunner-Munzel estimator \cite{Brunner2000}, $\hat{p} = P(\Phi_S < \Phi_N) + 0.5\,P(\Phi_S = \Phi_N)$, the probability that a value chosen at random from the northern values is greater than a value chosen at random from the southern values. It equals 0.5 for two equivalent hemispheres and makes no assumption on the shape or width of the distributions.

\subsection{Dispersion of the flux distributions and stability of the hemispheric ratios}
\label{sec:uncertainty}

The flux distributions presented in Section~\ref{sec:results} are strongly non-Gaussian and span more than two orders of magnitude in every sub-region and hemisphere. Table~\ref{tab:uncertainty} reports, for each sub-region, flux direction and hemisphere, the sample median together with the 16th and 84th percentiles of the underlying distribution. These percentiles measure the intrinsic spread of the sampled population, not the precision with which the center of that population is determined, and the two should not be confused, since two distributions whose percentile ranges largely overlap can still have medians that differ systematically, which is the situation encountered in the PE and the ME.

Successive JEDI spectra acquired during a single auroral crossing are separated by only $\sim$1\,s and sample the same magnetospheric state (source plasma population and solar wind driving), which is not expected to change significantly over the duration of the crossing, so they are strongly correlated. The effective number of independent realizations behind each median is therefore set by the number of contributing perijoves rather than by the number of grid cells, and a formal error bar based on the sample size would be misleadingly small. The northern and southern crossings of a given perijove are, in contrast, separated by a few hours, over which the magnetospheric state can change. This temporal variability adds scatter to the hemispheric comparison, which the subset test described below accounts for.
We therefore assess the robustness of the hemispheric ratios empirically, by recomputing them on subsets of the available perijoves. Random subsets drawn without replacement, from five perijoves upward, and the PJ3--PJ35 subset taken separately, return the same behavior of the ratio in the PE and in the ME, for both flux directions, and amplitudes scattered by a few tens of percent around the values quoted in Section~\ref{sec:results}. The reversal of the asymmetry between the two sub-regions is recovered from every subset tested. The hemispheric contrast reported here is therefore a stable property of the dataset and not an artifact of the particular set of perijoves analyzed.

\begin{table}[h]
\caption{Median downward and upward energy flux $\Phi$, in \mwm{}, with 16th and 84th percentile bounds ($\Phi_{16}$, $\Phi_{84}$), for each auroral sub-region and hemisphere. $n$ is the number of grid-cell averaged spectra contributing to each distribution (Table~\ref{tab:sample_sizes}). All values are given to three significant figures.}
\centering
\begin{tabular}{llcccc}
\hline
& & $n$ & Median & $\Phi_{16}$ & $\Phi_{84}$ \\
\hline
PE, down & N & 3819  & 5.03 & 1.01  & 37.1 \\
PE, down & S & 1005  & 2.05 & 0.67  & 15.7 \\
PE, up   & N & 5083  & 13.9 & 2.98  & 67.5 \\
PE, up   & S & 5441  & 6.99 & 2.33  & 30.5 \\
ME, down & N & 2123  & 5.85 & 1.88  & 24.6 \\
ME, down & S & 760   & 9.00 & 2.78  & 127  \\
ME, up   & N & 2024  & 4.84 & 1.46  & 25.0 \\
ME, up   & S & 802   & 9.62 & 2.31  & 113  \\
\hline
\end{tabular}
\label{tab:uncertainty}
\end{table}

Table~\ref{tab:ratio} reports, in the same way, the median ratio \ratio{} in each sub-region and hemisphere. Because this ratio requires both the downward and the upward flux in the same grid cell, $n$ is here the number of cells where both are available, and is therefore smaller than the corresponding counts of Table~\ref{tab:sample_sizes}.

\begin{table}[h]
\caption{Median ratio \ratio{} for each auroral sub-region and hemisphere, with 16th and 84th percentile bounds. $n$ is the number of grid cells for which both the downward and the upward flux are available.}
\centering
\begin{tabular}{llccc}
\hline
& & $n$ & Median & $[P_{16}; P_{84}]$ \\
\hline
PE & N & 2935 & 2.05 & [0.80; 7.05] \\
PE & S & 852  & 2.97 & [1.09; 14.4] \\
ME & N & 1406 & 0.72 & [0.40; 1.16] \\
ME & S & 311  & 0.72 & [0.47; 1.03] \\
OE & N & 7479 & 0.89 & [0.57; 1.11] \\
OE & S & 1645 & 0.96 & [0.78; 1.05] \\
\hline
\end{tabular}
\label{tab:ratio}
\end{table}

\section{Results and discussion}
\label{sec:results}

\begin{figure}
\centering
\includegraphics[width=\textwidth]{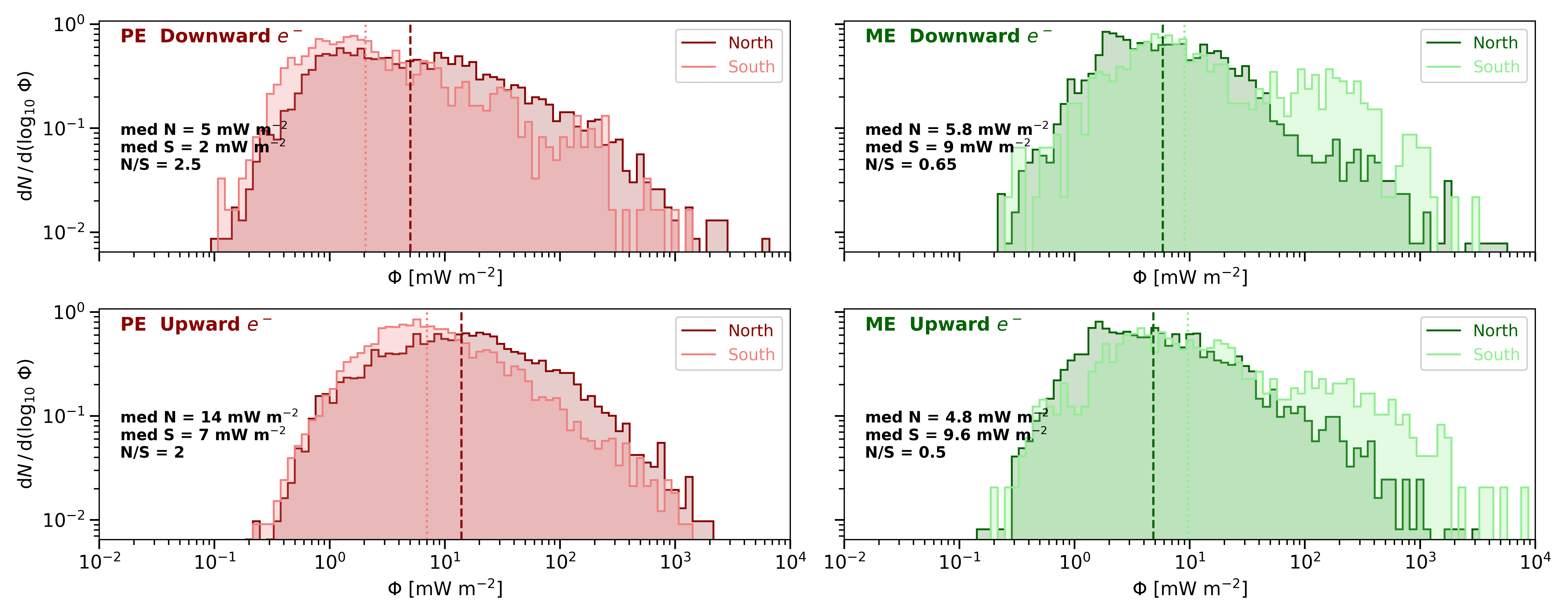}
\caption{Statistical distributions d$N$/d(log$_{10}\,\Phi$) of the downward (top row) and upward (bottom row) energy flux, in the polar (PE, left, red) and main (ME, right, green) emission regions. Dark and light lines correspond to the northern and southern hemispheres, vertical dashed and dotted lines to their respective medians, whose ratio is given as N/S. Note that the width of each distribution reflects the spread of the sampled population and not the precision of its median (Section~\ref{sec:uncertainty}). JEDI data, PJ3--PJ66.}
\label{fig:histos}
\end{figure}

\subsection{North-south asymmetry of the downward flux in the PE and ME regions}
\label{sec:asym}

Figure~\ref{fig:histos} (top row) presents the statistical distributions of the downward energy flux $\Phi_\downarrow$ in the PE and the ME regions, comparing the northern (dark line) and southern (light line) hemispheres. In the PE and ME the northern and southern $\Phi_\downarrow$ distributions broadly overlap, but each of them spans more than two orders of magnitude in $\Phi_\downarrow$, so their overlap does not by itself indicate agreement between the hemispheres. Their medians, however, differ by a factor of about two in both sub-regions, with N/S $=2.5$ in the PE and $0.65$ in the ME (Table~\ref{tab:uncertainty}). This contrast involves the whole $\Phi_\downarrow$ distributions and not only their centers, as a northern value exceeds a southern one in 62\% of cases in the PE ($\hat{p} = 0.62$), but in only 39\% of cases in the ME ($\hat{p} = 0.39$), a moderate but systematic dominance consistent with the broad overlap visible in Figure~\ref{fig:histos}. However, the asymmetry is reversed between the two sub-regions, the north dominating in the PE and the south in the ME. Which hemisphere dominates does not depend on the perijoves retained, and the amplitude of the ratios varies by only a few tens of percent, both being recovered from random subsets of as few as five perijoves and from the PJ3--PJ35 subset alone, so the asymmetry is a stable property of the dataset rather than a sampling fluctuation.

That the contrast reverses between the polar cap and the main oval means that the two hemispheres do not simply receive different amounts of energy, but partition it differently between sub-regions, which a single hemispheric scaling factor such as the internal field strength or the ionospheric conductance could not produce alone. Multiplying the median downward flux by the median area of each sub-region, derived from the UVS contours of each perijove \cite{Groulard2024}, gives a characteristic precipitated power above 30\,\keV{} of about 5.0 and 2.1\,TW in the northern PE and ME, against 1.8 and 3.5\,TW in the south, that is about 7.1 and 5.3\,TW for the two sub-regions combined. The hemispheric contrast in total power (N/S of about 1.3) is therefore much weaker than in each sub-region taken separately.

In the ME, where the precipitating flux is mainly associated with field-aligned acceleration producing inverted-V and broadband spectra \cite{Mauk2020,Salveter2022}, the southern excess matches, in both direction and order of magnitude, the factor by which southern field-aligned currents exceed northern ones \cite{Kotsiaros2019}. Whether the two are related cannot be settled here. A current is carried by the number flux of electrons at energies largely below the JEDI range, whereas we measure an energy flux above 30\,\keV{}, and \citeA{Kotsiaros2019} report a current integrated across the auroral zone rather than a local median. It is nonetheless, to our knowledge, the first time that such a hemispheric contrast is reported in the precipitating energy flux itself.

\subsection{The upward flux and the origin of the asymmetry}
\label{sec:upward}

The bottom row of Figure~\ref{fig:histos} presents the upward flux $\Phi_\uparrow$ distributions, built in the same way. The picture is unchanged. The same hemisphere dominates in each sub-region, with nearly the same amplitude (N/S $=2.0$ in the PE and $0.5$ in the ME, with $\hat{p} = 0.61$ and $0.38$). The asymmetry is therefore a property of the whole energetic electron population sampled in each sub-region, and not of its precipitating fraction alone.

The distributions of the mean electron energy $\langle E \rangle$ of each spectrum (not shown) peak at approximately the same energy in both hemispheres for every sub-region and flux direction, from $\sim$150--250\,\keV{} in the ME to $\sim$200--280\,\keV{} in the PE. A hemisphere in which electrons were accelerated more efficiently would display a higher characteristic energy, which is not observed. The factor-of-two contrast in energy flux therefore resides in the number flux rather than in the energy per particle, and points to a difference in the number of electrons reaching each hemisphere rather than in the efficiency of local acceleration. The number flux is also the quantity that carries the field-aligned current, which gives some physical substance to the comparison raised in Section~\ref{sec:asym}, without resolving it.

Comparing the two rows of Figure~\ref{fig:histos} rather than the two hemispheres, the upward and downward fluxes are of comparable magnitude in the ME. In the PE, by contrast, the upward distribution peaks at higher values in both hemispheres ($14$ against $5$ \mwm{} in the north, $7$ against $2$ \mwm{} in the south), so upward-going electrons carry the larger energy flux there.

\subsection{Directional anisotropy in the polar cap: mapping the up/down ratio}

\begin{figure}
\centering
\includegraphics[width=\textwidth]{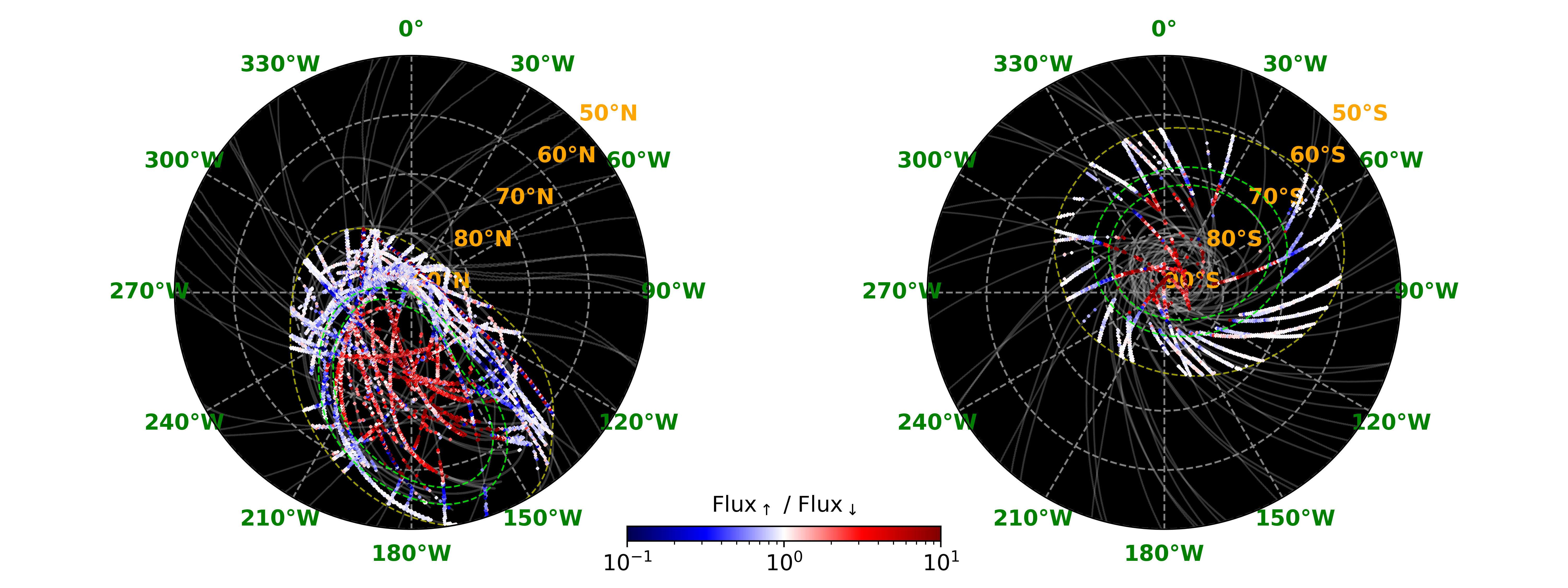}
\caption{Polar maps of the ratio \ratio{} in the northern (left) and southern (right) hemispheres, combining PJ3--PJ66, on a logarithmic color scale. Green dashed contours mark the inner and outer boundaries of the ME; the yellow dashed contour marks the mean magnetic footprint of Io, taken as the outer boundary of the OE. Gray lines show Juno's magnetic footprint trajectory for all perijoves; colored markers indicate the ratio only where a valid measurement falls within one of the three auroral sub-regions (PE, ME, OE).}
\label{fig:maps}
\end{figure}

Figure~\ref{fig:maps} presents the polar maps of the ratio \ratio{} in both hemispheres, combining all data from PJ3 to PJ66. In the ME, the ratio remains within a factor of about 1.4 of unity, with a median of $0.72$ in both hemispheres (Table~\ref{tab:ratio}), indicating a slight dominance of the downward flux consistent with the comparable upward and downward distributions of Figure~\ref{fig:histos}. The coverage is however markedly sparser than in the other sub-regions.

In the PE, by contrast, the upward/downward energy flux ratio greatly exceeds unity densely and consistently across the sub-region (median of $\sim2.0$ in the north and $\sim3.0$ in the south), reaching the upper end of the color scale (a factor of 10) near the magnetic pole and decreasing toward the main oval. This structure coincides with the swirl region \cite{Grodent2003b}, where the most energetic upward beams, reaching MeV energies and diffused in pitch angle by whistler-mode waves, have been associated with field-aligned stochastic acceleration producing broadband, power-law electron spectra \cite{Mauk2017_beams,Elliott2018a}.

The OE region is excluded from the energy flux statistics presented in Sections~\ref{sec:asym} and \ref{sec:upward}, because the measured flux there is increasingly contaminated by penetrating radiation from Jupiter's radiation belts as the magnetic footprint approaches the outer perimeter of the sub-region, near the mean magnetic footprint of Io (Section~\ref{sec:mapping}). The ratio \ratio{} remains partly informative there. Since the upward and downward intensities are derived from the same measurement at each time step, this penetrating background, which is not expected to favor either direction, adds to both of them. It therefore compresses the ratio toward unity without changing which direction dominates, so that the ratio still indicates whether the upward or the downward flux prevails, but no longer measures the degree of anisotropy. So computed, the ratio remains close to unity across the entire OE ($0.89$ in the north and $0.96$ in the south). Its median, slightly below unity, points at most to a weak dominance of the downward flux, and cannot distinguish a directionally isotropic population from a background-dominated signal. Near the outer boundary of the ME, where the contamination is weakest, this near-unity ratio is in good agreement with the measurements reported by \citeA{Mauk2020} (their Fig. 3), \citeA{Allegrini2020} (their Fig. 1) and \citeA{head_characterising_2026}.

Critically, the PE anisotropy is qualitatively identical in both hemispheres. It is therefore a common property of the Jovian polar cap rather than a consequence of the north-south asymmetry of the internal field. It also constrains the origin of the flux asymmetry established above. A comparable \ratio{} in two physically distinct polar cap environments (strongly non-dipolar in the north, predominantly dipolar in the south) indicates that the field-aligned acceleration operating in the PE is equally efficient in both hemispheres. The factor of two between the northern and southern PE fluxes therefore does not originate in the auroral acceleration itself, but in the number of electrons supplied to each hemisphere. This is consistent with the equality of the mean electron energies noted in Section~\ref{sec:upward}, and the two results together suggest that the asymmetry is set before the electrons reach the auroral acceleration region, for instance by the mapping of the common magnetodisk source onto each hemisphere or by the wave-particle interactions that scatter electrons into the loss cone.

\section{Conclusions}
\label{sec:conclusion}

We have carried out the first statistical study of the hemispheric asymmetry of energetic electron precipitation in the PE and ME auroral sub-regions of Jupiter, distinguishing the upward and downward fluxes, using JEDI data from PJ3 to PJ66. Our main results are the following.

\begin{enumerate}
\item In the PE and ME, the median electron energy flux differs between hemispheres by a factor of about two, for both flux directions, but the dominant hemisphere is reversed between the two sub-regions (median downward flux of $5.0/2.1$\,\mwm{} in the PE and $5.9/9.0$\,\mwm{} in the ME, median upward flux of $13.9/7.0$ and $4.8/9.6$\,\mwm{}, north/south). The two hemispheres therefore partition the precipitated energy differently between the polar cap and the main oval, a phenomenon that cannot be explained by any single hemispheric scaling factor.

\item In the PE, the upward flux systematically exceeds the downward flux in both hemispheres, with a radial structure peaking near the magnetic pole, consistent with stochastic field-aligned acceleration in the swirl region, whereas the ratio stays close to unity in the ME. Being qualitatively identical in the two hemispheres, this anisotropy indicates that the efficiency of field-aligned acceleration in the PE is the same in both. Together with the equality of the mean electron energies, this suggests that the flux asymmetry arises from the number of electrons reaching each hemisphere rather than from their acceleration. This could reflect differences in how the common magnetodisk source maps onto each hemisphere, or in the wave-particle interactions that scatter electrons into the loss cone.
\end{enumerate}

That the hemispheric asymmetry of the electron energy flux reverses between the PE and the ME, while the mean electron energy and the up/down anisotropy remain hemispherically symmetric, offers a new observational constraint for models of magnetosphere-ionosphere coupling, which have generally treated the two hemispheres as equivalent \cite{Bonfond2020}. Whether this contrast is connected to the comparable asymmetry reported in field-aligned currents \cite{Kotsiaros2019} remains open, and would require joint particle and magnetic field measurements to address it. These asymmetries also bear on the response of the auroral upper atmosphere. Since the mean electron energy is the same in both hemispheres, the precipitating electrons deposit their energy at comparable depths in the north and in the south, but with a downward energy flux about twice as large in the northern PE and in the southern ME. The heating and the chemistry driven by auroral electron precipitation \cite{Sinclair2018,Cavalie2023,Hue2024} should therefore not be modeled with a single precipitation input for both hemispheres, the northern polar cap and the southern main oval receiving the larger share of the energy above 30\,\keV{}. Only the downward flux contributes to this deposition, so the upward excess observed in the PE does not add to the energy delivered to the atmosphere there. Future work combining these JEDI results with systematic UV brightness maps from Juno/UVS will directly test the connection between in situ precipitation asymmetry and the morphology of the observed auroral emissions.

\section*{Open Research Section}

The JEDI level-3 electron data used in this study are publicly available from the NASA Planetary Data System \cite{Mauk2026}. The grid-cell averaged downward and upward energy fluxes displayed in Figures~\ref{fig:histos} and \ref{fig:maps} and reported in Tables~\ref{tab:sample_sizes} to \ref{tab:ratio} are archived on Zenodo \cite{Benmahi2026_zenodo}. Juno positions were computed with SpiceyPy from the Juno SPICE kernels, and figures were made with Matplotlib.

\section*{Conflict of Interest disclosure}
The authors declare there are no conflicts of interest for this manuscript.

\acknowledgments
B. Benmahi and V. Hue acknowledge support from the French government under the France 2030 investment plan, as part of the Initiative d'Excellence d'Aix-Marseille Universit\'e, A*MIDEX AMX-22-CPJ-04. French co-authors acknowledge the support of CNES for the Juno and Juice missions. B. Benmahi acknowledges support from a CNES postdoctoral fellowship. The authors thank B. Mauk and the JEDI team for helpful discussions on the JEDI calibration and the penetrating background.
B. Bonfond is a Research Associate of the Fonds de la Recherche Scientifique, FNRS. Authors acknowledge support from the JAFAR project (ANR-25-CE49-6683).

\bibliography{biblio}

\end{document}